\documentclass[12pt,titlepage]{article}  

\usepackage{graphicx}
\usepackage{amssymb,amsmath}
\usepackage{color}
\usepackage{caption}

\begin{document}

\begin{titlepage}
	\centering
    {\Large \textbf{Functional Brain Imaging: A Comprehensive Survey}\\}
    	\vspace{2cm}
    {\large Saman Sarraf$^1$, Jian Sun$^2$\\}
	\vspace{0.5cm}
	{\normalsize \textit{$^1$Electrical and Computer Engineering, McMaster University, Hamilton, Ontario, Canada} \\ 
	$^2$\textit{Department of Engineering Physics, McMaster University, Hamilton, Ontario, Canada}\\}
	\vspace{1cm}
	{\large \textbf{samansarraf@ieee.org}}
	\vfill
\end{titlepage}

\tableofcontents

\section{Introduction}

Functional brain imaging allows measuring dynamic functionality in all brain regions. It is broadly used in clinical cognitive neuroscience as, well as in research. It will allow the observation of neural activities in the brain simultaneously. 

From the beginning when functional brain imaging was initiated by the mapping of brain functions proposed by phrenologists, many scientists were asking why we need to image brain functionality since we have already structural information. Simply, their important question was including a great answer. Functional information of the human brain would definitely complement structural information, helping to have a better understanding of what is happening in the brain.

This paper, which could be useful to those who have an interest in functional brain imaging, such as engineers, will present a quick review of modalities used in functional brain imaging. We will concentrate on the most used techniques in functional imaging which are functional magnetic resonance imaging (\textit{fMRI}) and functional optical imaging, which is one of novelties in this area of study.

\section{Brain}

\subsection{Structure and Function}

The human brain is the most complex organ in the human body, having a complicated anatomy and physiology, which is working with almost all other organs through trillions of synapses \cite{goetz2007textbook}. Our brain includes many complicated areas, working based ganglion, and cerebellum. There are four lobes. They are frontal lobes, parietal lobes, temporal lobes, and finally the occipital lobes \cite{wong2006lucy}.

Generally speaking, the cortex is the outer most layers of brain cells which control voluntary movements and thinking. The brain stem, which is located between the spiral cord and the rest of the brain, controls our breathing and sleep. The frontal lobes manage problem solving, judgement and motor function while the parietal lobes manage body position, hand writing, and sensation. The temporal lobes are responsible of memory and hearing, and the occipital lobes contains the brain’s visual processing system \cite{goetz2007textbook,wong2006lucy}.

\subsection{Neurovascular Coupling}

Crucial to interpreting functional imaging data and normal brain functions, is to have a deep understanding of neurovascular coupling (Toga, 2012). There is a close relationship between spatial and temporal coupling and neuronal activity, as well as blood flow that ensures all brain areas are using a proper amount of oxygen and receiving energetic metabolites. The most recent research shows the activation of Ca$^{2+}$ elevations is a necessary step in neuronal activities \cite{carmignoto2010contribution}.

As is known, neurons have a tight relationship with astrocytes, smooth muscle, pericytes, endothelial cells, and erythrocytes. Neuronal chemo-ele\-ctrical activity is speculated to be associated with many metabolic procedures, which are called neurovascular coupling. Neurovascular coupling includes the following steps: Glucose is metabolized to lactate in astrocytes, and then lactate is travels to neurons and metabolized with oxygen to produce ATP and carbon dioxide. Arteries are responsible to deliver oxyhemoglobin to neurons and in the presence of carbon dioxide, and oxygen will be released. Consequently, Hbo is converted into dHbo. Neurotransmitters such as acetylcholine released by active neurons cause relaxation step in smooth muscles, and in arterioles, which will increase blood flow and volume \cite{toga2002brain}.

\begin{figure}[h]
\centering
\captionsetup{justification=centering,margin=0.5cm}
\includegraphics[width=0.95\textwidth]{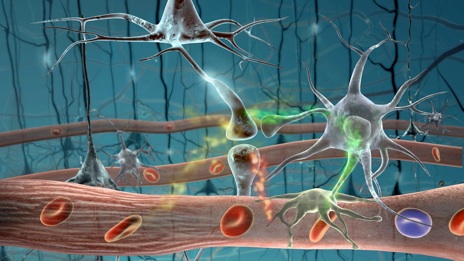}
\caption{Neurovascular coupling \cite{toga2002brain}.}
\label{fig101}
\end{figure}

\subsection{Temporal and spatial resolution}

A novel field of study combining cognitive science and cognitive psychology as well as biology and neuroscience has been developed and called cognitive neuroscience \cite{ward2015student}.

In 1988, Churchland and Sejnowski proposed a method of brain function study based on temporal and spatial resolution of modalities used in this area. They supposed that it could be possible to categorize functional brain imaging modalities into temporal resolution and spatial resolution. This model has been developed several times, and now many novel modalities have been created. Briefly, temporal resolution refers to the accuracy measured, when an event is happening, and spatial resolution refers to the accuracy measured, where an event is occurring \cite{ward2015student}.

Each modality is categorized by the ability to define brain functionality based on detectable range in space and time. Modalities located near the bottom of Figure 2 have better spatial resolution than those at the top while methods situated at the left of figure show better temporal resolution. Therefore, modalities such as EEG, MEG and TMS demonstrate a good temporal resolution while functional MRI and PET, have seconds and minutes temporal resolution \cite{lane2009rebirth}.

As is known, functional MRI and functional optical imaging - which has become more popular today - are the most used imaging techniques in the functional brain imaging. In next part, we will look at modalities and concentrate on \textit{fMRI} and \textit{fNIRS}.

\begin{figure}[h]
\centering
\captionsetup{justification=centering,margin=0.5cm}
\includegraphics[width=0.95\textwidth]{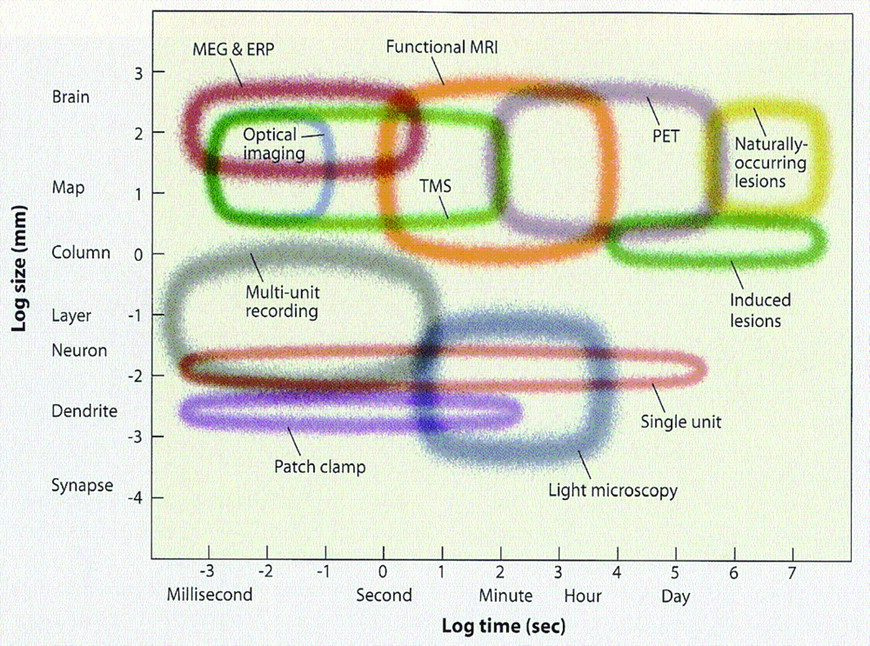}
\caption{Spatial Vs. Temporal Resolution: Gazzaniga,1998.}
\label{fig102}
\end{figure}

\section{Imaging Modalities}

\subsection{Qualitative EEG and MEG}

Electroencephalography of EEG is a technique for the recording of electrical signals coming from brain activities. As is known, an EEG recording is based on voltage fluctuations resulting from ionic current flows within the neurons in the brain. As mentioned before, EEG has a good temporal resolution and these days it is used in dual modality fMRI-EEG in the functional brain \cite{niedermeyer2005electroencephalography}.

Also, Magnetoencephalography is a technique for mapping activity. The electrical currents which are naturally happening in the brain produce magnetic fields which are recorded by MEG technique. Application of MEG includes basic research in cognitive brain processes etc. \cite{lauronen2012magnetoencephalography}.

\subsection{Position emission tomography (PET)}

PET presents a nuclear medicine imaging technique producing 2-D and 3-D images of brain functionality. Physical principles of PET imaging include several steps which begin with choosing the molecule of interest. This molecule has to be located radioactively and the product of this step is called tracer. PET imaging is based on detecting pairs of gamma rays which are indirectly emitted by the positron-emitting radionuclide (tracer) - which will play the role of an active molecule in the body. When sufficient amounts of the labelled molecule are administered into the body, the patients carrying the radioactive trace are placed into the PET scanner. The basic of image acquisition is similar to all computed-tomography techniques. The radiation caused by the annihilation step will be detectable in the output. The next step is image reconstruction. The final goal of PET imaging is to measure the distribution of the tracer. Images obtained by PET imaging will be filtered by cut-off values, helping the radiologist to interpret the results \cite{valk2006positron}.

The application of PET is not limited to neuroimaging and it can be used on oncology, cardiology, musculo-skeletal and small animal imaging. The most popular active molecule in PET imaging is Fludeoxyglucose (18F) called FDG which is an analogue of glucose \cite{kiendys2009f}.

\subsection{Single--Photon Emission Computed Tomography (SPECT)}

Brain SPECT is a powerful clinical and research tool in the management and diagnosis of neurologic diseases \cite{morano2003technical}.

As is known, SPECT is a nuclear medicine imaging modality providing information of cerebral function and like other nuclear medicine technique, using a radio pharmaceutical tracer \cite{harrison1999single} \cite{sarraf2009simulation} \cite{sarraf2014brain} \cite{sarraf2014mathematical}. The poor image quality causes less usage in clinics. The procedure in SPECT is similar to PET but the major difference is that radionuclide used in PET is emitting two gamma-photons while there is only one produced photons in SPECT. Therefore, the image quality of PET is much higher than SPECT. This modality is utilized in functional brain imaging and also in myocardial perfusion imaging.

\subsection{Functional Magnetic Resonance (\textit{fMRI})}

\subsubsection{Introduction}

\textit{fMRI} is a technique measuring the brain activity by detecting the associated changes in blood flow. This technique, which is similar to MRI, uses the change in magnetization between oxygen-rich and oxygen-poor blood happening in the brain areas as its basic measure. 

The idea for foundations of magnetic resonance imaging were almost started in 1946 by Felix Bloch (1905-1983) who was at Stanford University studying liquids, and Edward Purcell (1912-1997) working on solids at Harvard University. Although they could have won Nobel Prizes for these discovers, it was not until 1973, that successful nuclear magnetic resonance (NMR) was used to produce images. Finally, in the 1990s their discovers showing changes in blood oxygenation level could be considered with the use of MRI as a window opening a new generation of functional brain imaging techniques \cite{huettel2004functional} \cite{strother2014hierarchy} \cite{grady2016age}.

\subsubsection{MR Signal Generation and Formation}

A variety of physical concepts resulted in magnetic resonance signal generation. First, an atomic nuclei having magnetic momentum as well as angular momentum which are called nuclear spin, plays the important role. ``Two axes are important here which are longitudinal direction and transverse plane, respectively. Longitudinal direction is the axis around which an atomic nucleus precesses and the transverse plane is the plane on which a nucleus precesses.'' \cite{huettel2004functional}.

As is known, each spin will receive a low or high energy state (different magnetic potential energies) which could be parallel or antiparallel to the magnetic field. By applying an electro-magnetic pulse oscillating at the resonant (Lamor) frequency of the spins and this procedure is called excitation; the net magnetization vector will change from longitudinal into the transverse plane \cite{huettel2004functional}. The mentioned step called evolution will make the net magnetization to change over time in transverse plane which will be generating MR signals. MR signals can be measured in the external coil. The Bloch equation describes quantitatively the behaviour of magnetization in the presence of a time-varying magnetic field \cite{huettel2004functional,toga2002brain}.

\begin{center}
{\color{blue} Equation 1 the Bloch Equation}
\end{center}
\begin{equation*}
\frac{dM}{dt}=\gamma M \times B+\frac{1}{T_1}(M_0-M_z)-\frac{1}{T_2}(M_x+M_y)
\end{equation*}
\begin{center}
{\color{blue} The Bloch Equation, a quantitative description of magnetic resonance where $M$ is net magnetization, $\gamma$ is gyromagnetic ration, $B$ is magnetic field, $T_1$ is the time constant of longitudinal relaxation process and $T_2$ is the time constant of transverse relaxation.}
\end{center}

As described by the Bloch equation, the net magnetization of a spin can be divided into separate special components along the spatial coordinates $(x,y,z)$. ``The longitudinal magnetization is defined as $M_z$ and the transverse magnetization are defined as $M_{xy}$'' \cite{huettel2004functional}.

$T_1$ contrast is related to the time of recovery of the longitudinal magnetization (excitation) and $T_2$ contrast is associated with the decay of transverse magnetization \cite{huettel2004functional}. Briefly, there are several steps to achieve a 3-D MR image. First, as mentioned before, longitudinal magnetization happens following the volume excitation. The next step is transverse magnetization, where 3-D spatial encoding is performed in order to acquire a MR signal in K-space, resulting in a 3-D MR image \cite{huettel2004functional,toga2002brain}. 

K-space is a 3D Fourier transform of the MR image measured. Its complex values are sampled during an MR measurement, in a premeditated scheme controlled by a pulse sequence, i.e. an accurately timed sequence of radiofrequency and gradient pulses. In practice, k-space often refers to the temporary image space, usually a matrix, in which data from digitized MR signals are stored during data acquisition. When k-space is full (at the end of the scan) the data are mathematically processed to produce a final image. Thus k-space holds raw data before reconstruction. As described, image acquisition in MRI is using the K-space idea affected by a Fourier transform of the image space. 

\subsubsection{MR Contrast Mechanisms}

Generally, the mechanisms of MR contrasts are divided to four main categories. 1) Static Contrast which are sensitive to the type, number and relaxation properties of atomic nuclei. This category includes, $T_1$-weighted, $T_2$-weighted, $T_2^*$ weighted and proton density contrast. The main application of static contrast is to determine anatomic information of the brain in \textit{fMRI}. 2) Motion Contrasts are sensitive to the movement of spins in the space. This group which includes diffusion and perfusion-weighted, will define dynamic properties of the proton in the brain such as blood flow through MR angiography. 3) Endogenous contrast is related to an internal feature of biological tissue in the brain while 4) Exogenous contrast needs an external agent injected to the body playing the role of contrast. For instance, BOLD \textit{fMRI} is an endogenous contrast because it uses the blood flow properties, but Gadolinium (Gd) will be an exogenous contrast \cite{huettel2004functional}.

\begin{figure}[h]
\centering
\captionsetup{justification=centering,margin=0.5cm}
\includegraphics[width=0.8\textwidth]{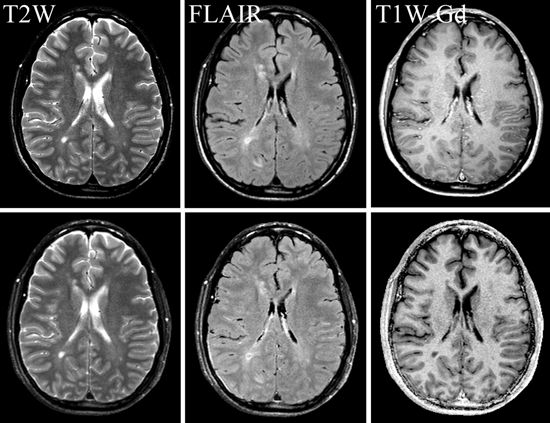}
\caption{Comparison of conventional and Synthetic Contrast MRI. $T_2$-weighted, $T_2$-weighted FLAIR and $T_1$-weighted post-Gd contrast of the same slice. The first row was acquired conventionally, the second row has been synthesized \cite{warntjes2008rapid}.}
\label{fig103}
\end{figure}

\subsubsection{Image contrast}

Basically, the number of spins in a sample volume will determine a MR signal resulting image contrast. The equation below shows that image contrast is affected by other parameters. In fact, image contrast will be determined by $T_R$ and $T_E$. For instance choosing a quite small $T_E$ will eliminate $T_2$-weighted images \cite{toga2002brain}. This short description might be of some mathematical interests.

\begin{center}
{\color{blue} Equation 2 Image Contrast Equation}
\end{center}
\begin{equation*}
S\sim\rho \frac{1-\exp\left( -\frac{T_R}{T_1} \right)}{1-\cos\alpha\exp\left( -\frac{T_R}{T_1} \right)}\sin\alpha\exp\left( -\frac{T_E}{T_2} \right)
\end{equation*}

Image contrast equation where $\rho$ is the density of nuclear spins, $T_R$ is repetition time, $T_E$ is echo time, and $\alpha$ is flip angle. This equation shows that $T_1$ will be maximized by $T_R$ and by using a flip angle equal or greater than the Ernest angle.

\subsubsection{BOLD \textit{fMRI}}

Blood Oxygenation Level Dependent \textit{fMRI} (BOLD \textit{fMRI}) is a technique detecting the brain and neural activity indirectly, based on magnetic properties of hemoglobin. It means that when Hbo and dHbo are placed in a magnetic field, dHbo will act as a contrast agent. In BOLD \textit{fMRI}, the change in the concentration of dHbo is dynamically monitored. Changes and correlations between tasks (stimulation) and concentration of Hbo will be considered. As a general fact, the more neural activities there are, the more tissue needs energy. This will increase the consumption of Glucose and O2, increasing the blood flow and volume, while local concentration of dHbo is decreasing which will produce the BOLD \textit{fMRI} signals. This process will be repeated to the end and it should mention that BOLD \textit{fMRI} signals are relevant to these three physiological concepts: Glucose and O2 consumption, blood flow and volume, and concentration of dHbo, respectively \cite{kim1997functional}.

\subsubsection{Pre-processing and Statistical Analysis of \textit{fMRI} data}

Preprocessing procedures measuring (removing) unexpected events is done in order to improve experimental analysis. Regular quality which is important for diagnosis problems has to be assured in this step. Preprocessing of \textit{fMRI} data are based on four principal phases. First, recording of \textit{fMRI} signal is done followed by an image reconstruction. Then the online quality assurance is applied. Quality assurance, called Q.A, is a set of procedures defining errors in \textit{fMRI} data, to avoid continuing the process with those errors. The second phase will be head motion correction, as well as slice timing correction. Each slice is acquired at a different time point within the TR, and the timing difference are especially problematic for an interleave sequence, in which spatially consecutive slices are not achieved successively. Consequently, it seems that having a slice timing correction step could be necessary. This phase is a recursive procedure and it is repeated to the best possible correction. In the third phase, distortion correction, functional and structural co-registration normalization as well as spatial and temporal smoothing (image smoothing) are respectively performed. Finally, the online quality assurance is repeated again and \textit{fMRI} image are provided \cite{huettel2004functional}.

Since the \textit{fMRI} provides vast information which could be difficult to interpret, a variety of statistical approaches are used in the analysis of the \textit{fMRI} data. The basin statistical tests such as t-test or K-S test, as well as frequency domain analysis methods such as Fourier or Wavelet analysis are applied on data of \textit{fMRI}. Furthermore, the pattern classification methods such as artificial neural networks or graph theory could be useful to study \textit{fMRI} data \cite{huettel2004functional,toga2002brain}.

\begin{figure}[h]
\centering
\captionsetup{justification=centering,margin=0.5cm}
\includegraphics[width=0.95\textwidth]{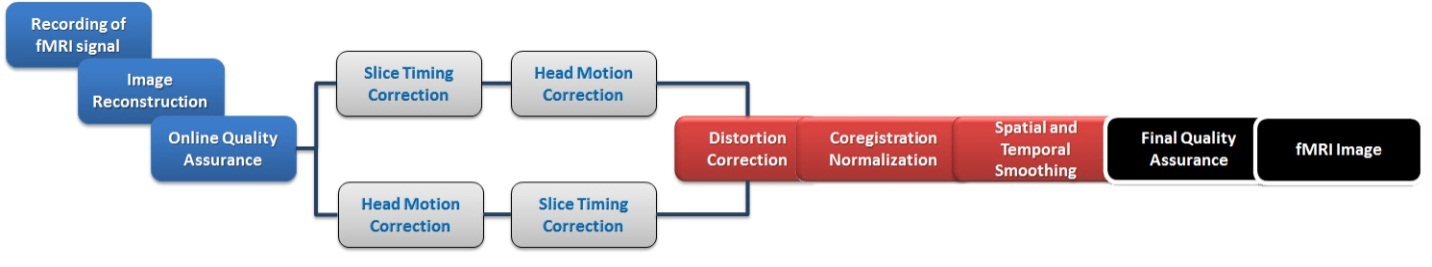}
\caption{Framework above shows the different phases of pre-processing of \textit{fMRI} data.}
\label{fig104}
\end{figure}

\subsubsection{Resting-state \textit{fMRI}}

A novel technique used in \textit{fMRI} called resting-state and as its name shows, the \textit{fMRI} data acquisition is performed while there is no task to do for patients. There are two related concepts about resting-state \textit{fMRI}. First of all, as patients do not need to do any task and there is no simulation, the procedure will be more comfortable than a normal \textit{fMRI} and secondly, as is known, \textit{fMRI} data interpretation is often difficult and it is a complicated work. Because in \textit{fMRI}, we are not sure of how much the task done by patients, could affect other brain regions or other cortex. Hopefully, in resting-state \textit{fMRI} it is supposed that the relationship between brain areas is in a normal condition and there is no specific effect from one cortex to another region \cite{bellec2010multi,liu2008regional}.

For studying resting-state \textit{fMRI} data, several methods can be applied. The most used approach is seed voxel. The related idea is that a voxel is considered in the MR image which is anatomically known for scientist. As each voxel represents a signal, they try to find all signals having high correlation with the seed voxel signal. After finding all correlated signals, they consider those in a cluster and a network in the brain is achieved \cite{bellec2010multi,liu2008regional}. 

\begin{figure}[h]
\centering
\captionsetup{justification=centering,margin=0.5cm}
\includegraphics[width=0.8\textwidth]{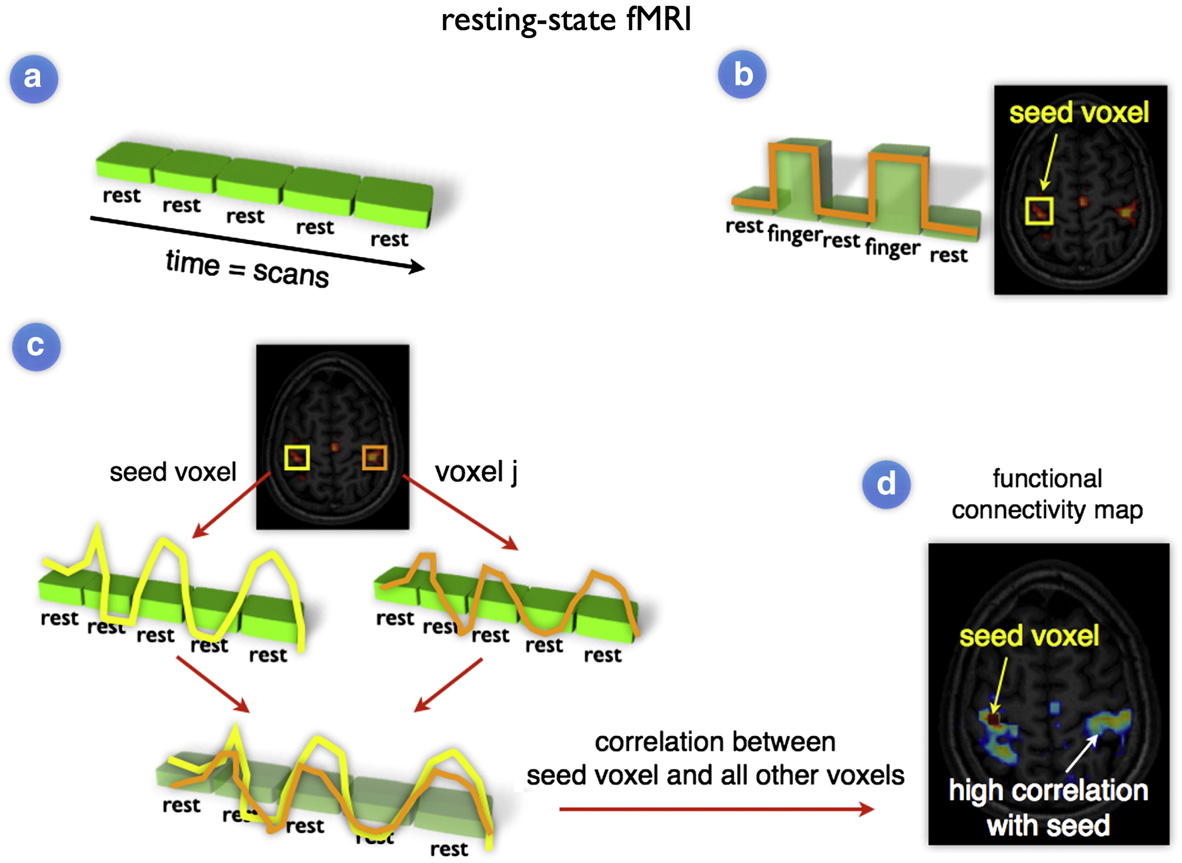}
\caption{Resting-state fMRI studies are focused on measuring the correlation between spontaneous activation patterns of brain regions \cite{van2010exploring}.}
\label{fig105}
\end{figure}

Another approach which is useful to study resting-state \textit{fMRI} data is called graph analysis. A graph including its clustering-coefficient and connectivity degree, etc. is formed and with the help of this graph, the networks in the brain could be found. According to \cite{van2010exploring} several networks such as primary motor, primary visual, extra-striate visual and insular-temporal/ACC have been found by using the above mentioned approaches. 

To summarize, ``Functional connections of resting-state networks is related to structural white matter connections suggesting a structural core of functional connectivity networks in the brain'' \cite{van2010exploring}. 

\begin{figure}[h]
\centering
\captionsetup{justification=centering,margin=0.5cm}
\begin{minipage}[b]{0.5\textwidth}
\centering
\includegraphics[width=0.95\textwidth]{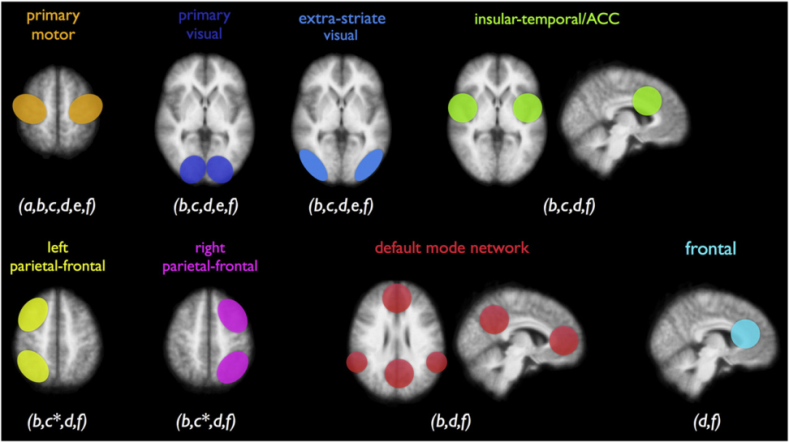}
\end{minipage}%
\begin{minipage}[b]{0.5\textwidth}
\centering
\includegraphics[width=0.95\textwidth]{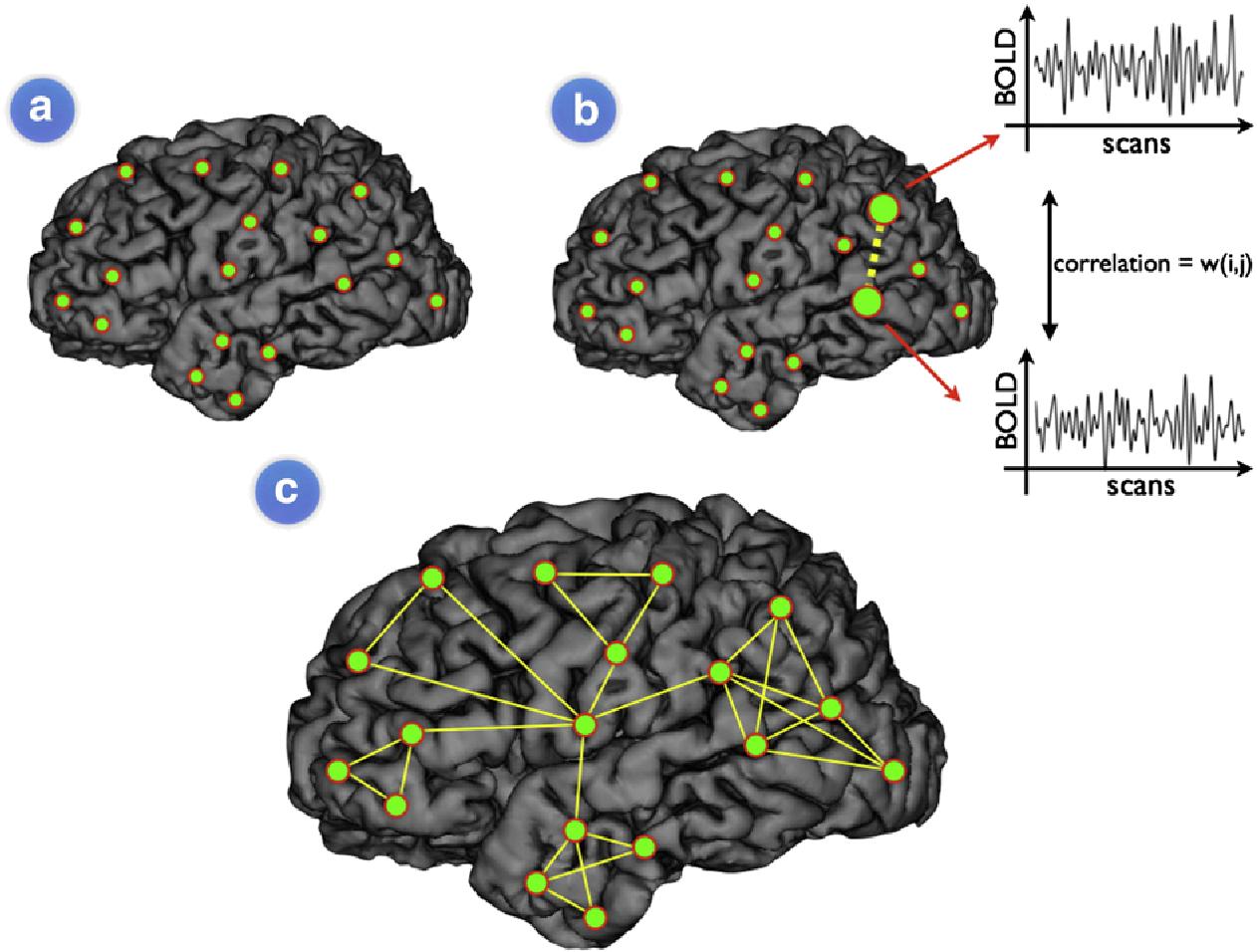}
\end{minipage}
\caption{Resting-state networks. A number of group resting-state studies have consistently reported the formation of functionally \cite{van2010exploring}.}
\label{fig106}
\end{figure}

\subsection{Functional Brain Imaging and Multimodalities}

Medical imaging by single modality could provide information needed for a diagnosis but the concept of the combination of two modalities in order to have more information, have been practically done from the past. For instance, although a patient is imaged by different modalities in different places, the diagnosis is performed by studying of all results. Thus, the idea of dual modality (hybrid modality) could be realized, especially when a combination of structural and functional imaging data were needed to lead to a better interpretation \cite{townsend2008dual}.

Several problems exist which avoids the multi-modality imaging system to be performed ideally. The major problem is the difference between temporal resolutions of modalities which has to be combined. Other software limitations such as image fusion and hardware restrictions are still a challenge in this area. Hopefully, lots of experiences have been done in dual modality and some of them are commercialized nowadays. Table 1 illustrates several imaging (non-imaging) systems used in hybrid modalities \cite{fazli2012enhanced}.

\begin{center}
\begin{minipage}{\textwidth}
\centering
{\color{blue} Table 1 Technical feasibility of simultaneous multimodal imaging \cite{fazli2012enhanced}.}
\includegraphics[width=0.95\textwidth]{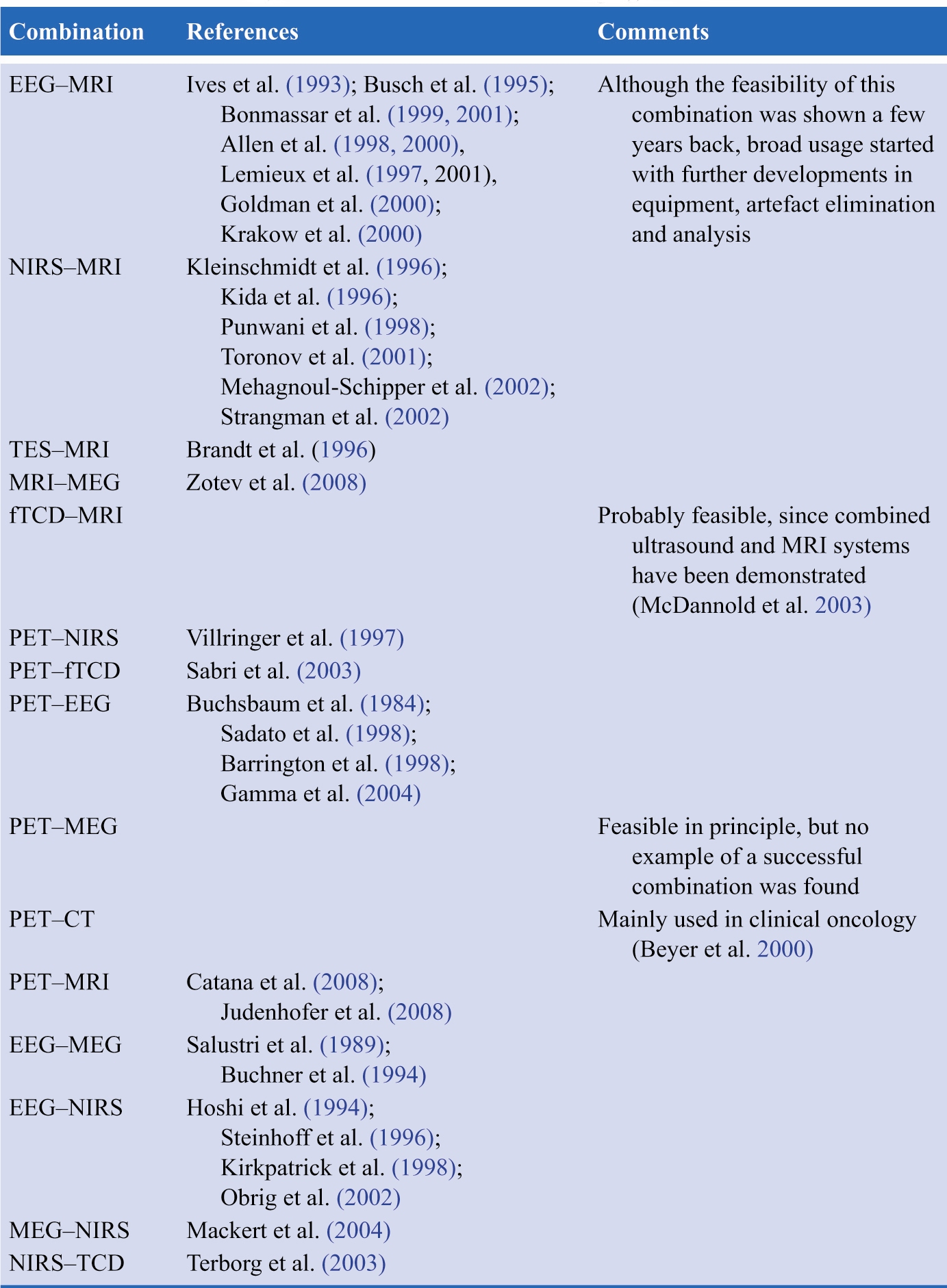}
\end{minipage}
\end{center}

\section{Functional Brain Imaging -- Optical Approach}

\subsection{Introduction}

It has been a long history since optical method was first used to capture functional activities in brain \cite{jobsis1977noninvasive,tsien1981non}. After thirty years of development, now it is able to perform non-invasive, three dimensional functional optical imaging of human brain or in vivo in living animal brain tissues. Functional optical brain imaging carries significant importance for understanding the physiological basis of neuron activities, because of its high temporal resolution and high sensitivity that other well-established clinical imaging modalities cannot provide. PET can provide 3-D image using fluorodeoxyglucose (FDG) labeled with radioactive isotope. However both spatial and temporal resolution is low. CT is able to provide high resolution image with fast data acquisition, but typically it cannot obtain functional information of the tissue. Functional MRI (fMRI) using the blood oxygen level dependent (BOLD) signal has been through significant development in recent years. However, the achievable temporal resolution is still limited by its long data acquisition time. Besides, optical imaging does not depend on ionizing radiation, has substantially lower infrastructure costs, and require minimum footprint and maintenance. 

Functional optical brain imaging can be either invasive or non-invasive. High resolution brain imaging requires direct exposure to laser light, such as multi-photon fluorescent microscopy \cite{denk1990two}, which has been widely adopted in animal neuron imaging researches. Non-invasive imaging generally refers to functional near-infrared spectroscopy (fNIRS), which detects the hemodynamics in brain tissues. It has provided useful complimentary information to clinical studies such as cognitive activity \cite{arenth2007applications} , Alzheimer's disease \cite{ferrari2012brief,hock1996near,irani2007functional}, stroke \cite{strangman2006near,nelson2006development} and birth asphyxia \cite{wyatt1989cerebral,van1993changes,toet2009brain} etc. The major problem for 3-D optical imaging is scattering, which has limited the achievable image quality and the penetrable depth. As a result, most optical brain imaging studies have been limited in the brain cortex.

\subsection{Microscopic Functional Imaging}

Microscopic study of neuron networks provides basic yet significant understanding of cellular and molecular activities in the brain. Analyzing the neural circuits in vivo helps develop network topologies that give us fundamental information about their interactions and functionalities. Two-photon microscopy has caught much attention since it was first invented in 1990 \cite{denk1990two}, a few years after the invention of Ti-sapphire laser in the early \cite{moulton1986spectroscopic} . Instead of exciting the fluorophores at visible wavelength, the two photon microscope utilizes femtosecond laser at near infrared wavelength to complete excitation via two photon absorption, which is a third-order nonlinear optical process. As a result, only the fluorophores at the focal point of the beam can be excited. 3-D scanning can be realized by adding galvo mirrors or acoustic optical deflectors (AODs) in the optical assembly. Since the infrared light can penetrate into tissues with lower scattering, and its wavelength is well apart from the fluorescent wavelength, the two photon microscope is able to provide 3-D image with unprecedented high spatial resolution. Figure 1 shows the schematic structure of the microscope. Over the past decade, two-photon microscopy has become a well-established method for in vivo optical neuroimaging \cite{jobsis1977noninvasive,ohki2006highly,zonta2003neuron,takano2006astrocyte} .

\begin{figure}[h]
\centering
\captionsetup{justification=centering,margin=0.5cm}
\includegraphics[width=0.7\textwidth]{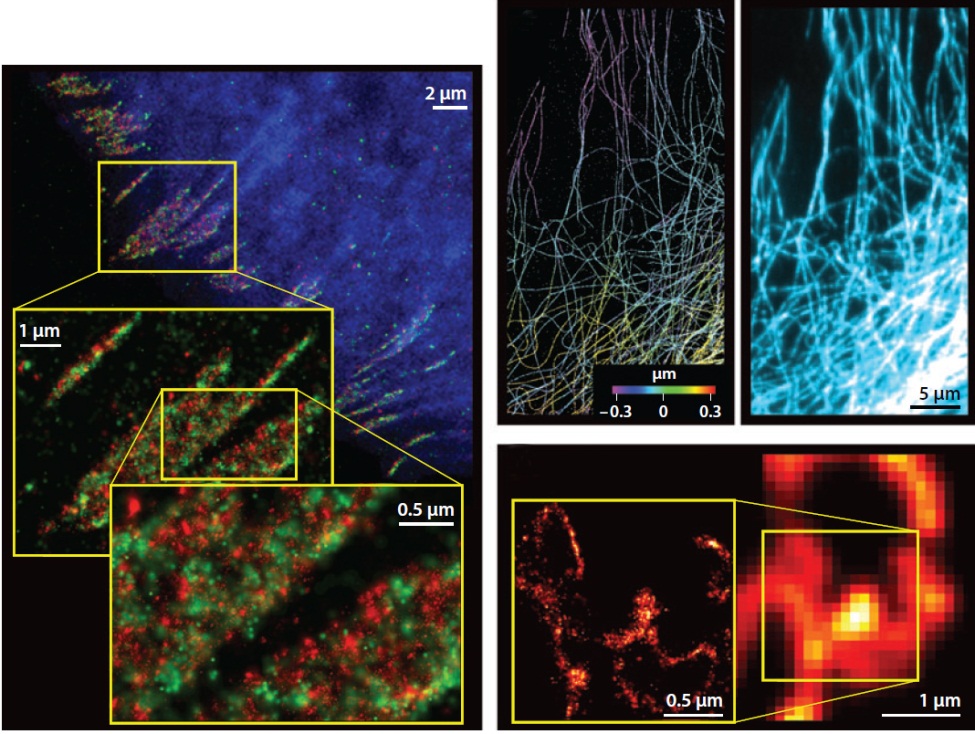}
\caption{Super-resolution fluorescent microscopy \cite{wilt2009advances}.}
\label{fig107}
\end{figure}

\begin{figure}[h]
\centering
\captionsetup{justification=centering,margin=0.5cm}
\includegraphics[width=0.9\textwidth]{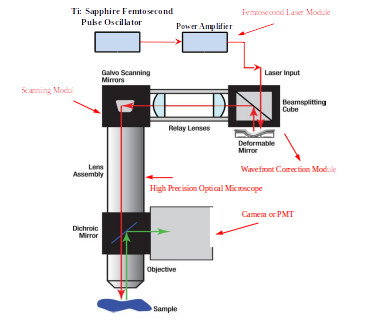}
\caption{Schematic diagram of a two photon fluorescent scanning microscope.}
\label{fig108}
\end{figure}

Intracellular calcium ion concentration is closely linked to many cellular functions. Functional imaging in vivo can be achieved by use of membrane-permeable calcium sensitive dyes such as Oregon Green 488 BAPTA-1 acetoxymethyl ester \cite{ohki2006highly,cossart2005calcium} . The dye is injected via a pipette, and delivered into interested area. After diffusing into the cell, it is hydrolyzed by endogenous esterase and stays inside the cell, while the remaining extracellular dye is gradually washed out. It results in a uniform staining of all cells in the field of view. This method is called the multi-cell bolus loading (MCBL) technique, by which means fluorescent observations can be performed in intact cortical neuron cells. An example of two photon microscope image of neuron ensembles is shown in Figure 7.

\begin{figure}[!h]
\centering
\captionsetup{justification=centering,margin=0.1cm}
\includegraphics[width=0.85\textwidth]{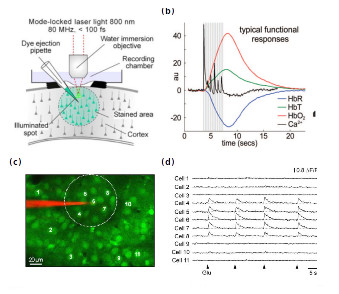}
\caption{(a) The MCBL method for neuron cell staining \cite{garaschuk2006targeted}(b) Typical Ca$^{2+}$ and hemodynamic response of a neuron cell after stimulation \cite{hillman2007optical}. (c) and (d). In vivo Ca$^{2+}$ transients in mouse visual cortex, the circle in (c) marks the area of glutamate stimuli.}
\label{fig109}
\end{figure}

In order to capture fast dynamics and large area imaging, fast scanning technique is applied. One of the most interested techniques is by employing AOD pairs \cite{bullen1997high}. By propagating sound waves through an optical crystal, AOD works as a tunable light deflector, and by changing the frequency of the acoustic waves, it steers the laser beam in one dimension. By orthogonally stacking multiple AODs, it can provide 3-D scanning abilities \cite{reddy2008three}. Comparing with mechanical scanning components such as galvo mirrors, AOD scanner has higher operating frequency, and can provide random access to the spot of interest. 

\begin{figure}[h]
\centering
\captionsetup{justification=centering,margin=0.5cm}
\includegraphics[width=0.95\textwidth]{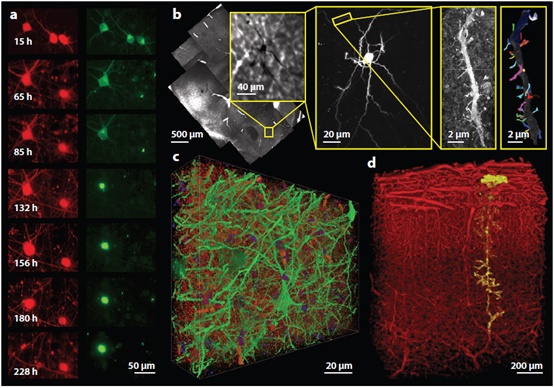}
\caption{Automated and multiscale fluorescent microscopy \cite{wilt2009advances} (a) Automated microscopy. (b). Multiscale microscopy. (c). Array tomography. (d). All-optical histology.}
\label{fig110}
\end{figure}

As shown in Figure 8, by employing the state-of-the-art technology, multiscale automated scanning microscopy can be realized in a large area and over several days of continuous observation.

Researches have been done in disease models by two photon calcium imaging, such as Alzheimer's disease (AD), stroke, ischemia, hypoxia and epilepsy \cite{rochefort2008calcium}. Although the cause and progression of AD remains not well understood, it is believed that the senile plaques and neurofibrillary tangles are associated with the disease \cite{tiraboschi2004importance}. The beta amyloid (A$\mathrm{\beta}$) can be fluorescent labeled in vivo, and by imaging A$\mathrm{\beta}$ and the neurons simultaneously, it is possible to study the toxicity of A$\mathrm{\beta}$ plaque and the evolution of neuronal damage \cite{meyer2008rapid}. Researches show that glial cells are also strongly involved in the development of AD. In vivo imaging of calcium signaling has shown abnormal astrocyte activity and instability at the early stage of the disease, before the onset of A$\mathrm{\beta}$ plaque deposition \cite{meyer2008rapid}. 

\subsection{Non-invasive Functional Imaging}

Hemoglobin is the most significant contributor to light absorption in brain. Since oxy- and deoxyhemoglobin (HbO$_2$ and HbR) possess unique and distinctive absorption spectra at visible and near infrared wavelength (Figure 4), its dynamics can be measured by spectroscopy. Functional activities in brain are closely correlated with cerebral blood flow, blood volume and oxygenation. As a result, brain functions can be captured by analyzing the spectra of hemodynamic response. 

Infrared light is preferable for spectroscopy because of its lower scattering rate in the tissue. And to determine HbO$_2$, HbR and Hb total (HbT) signal respectively, two or more wavelengths must be used simultaneously. Calculation of HbO$_2$, HbR and HbT concentration can be done by applying the Beer-Lambert law, while the effect of light scattering must be taken into consideration \cite{kohl2000physical}. Because of the scattering, photons have traveled different distances before collected by the detector. Most photons traveled through very superficial layers of the cortex, while few of them has penetrated deeply into the brain tissue and returned. The differential pathlength factor (DPF) is used to describe this differential attenuation effect. Light with different wavelengths possess different DPF, and it should impact the accuracy of the result when multiple wavelengths are used.

Comparing with other functional imaging modalities, fNIRS has demonstrated its unique usefulness. EEG has high temporal resolution, but it is difficult to determine the spatial location of the signal source. PET and fMRI provides excellent spatial resolution, but the data acquisition time is long. fNIRS is somewhat in between. It can acquire data rapidly with hundreds of Hertz frequency, and generate three dimensional spatial pictures combined with tomosynthesis, although without the ability to provide anatomical structures. For infants, PET and fMRI are not well suited because of the radiation risk and artifact cause by patient movement. Infrared light can penetrate through the skin and skull of a baby much easier than an adult, which also makes fNIRS an ideal method for functional brain imaging of awake infants (Figure 9). In addition, fNIRS is able to record brain reactions to auditory stimuli, which is impossible in fMRI.

\begin{figure}[h]
\centering
\captionsetup{justification=centering,margin=0.5cm}
\includegraphics[width=0.95\textwidth]{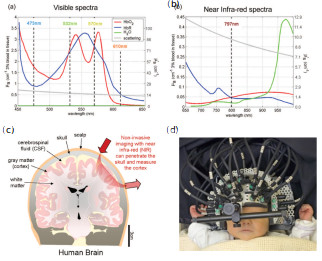}
\caption{Absorption spectra of HbO$_2$, HbR and HbT at (a) visible and (b) near infrared wavelength. Light scattering and water absorption curves are also shown. (c). Schematic of non-invasive fNIRS functional brain imaging and its application on an infant (d) \cite{hillman2007optical}.}
\label{fig111}
\end{figure}

\begin{figure}[h]
\centering
\captionsetup{justification=centering,margin=0.5cm}
\includegraphics[width=0.95\textwidth]{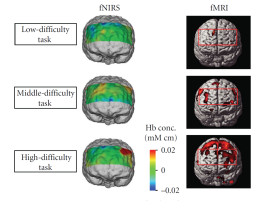}
\caption{Co-registered functional brain imaging by fMRI and fNIRS, showing good signal correlation \cite{tsunashima2009measurement}.}
\label{fig112}
\end{figure}

Although both BOLD fMRI and fNIRS are designated to detect hemodynamic signals, researches have revealed certain discrepancies. One most debated phenomenon is the ``initial dip'' \cite{malonek1996interactions}. Local HbR concentration increases in a short time after stimulation, before it enters a substantial decrease. One possible explanation for this transient change is that local HbO$_2$ is converted to HbR quickly after activation, before it is replenished from blood flow. However, this dip is only seen in a few BOLD fMRI and many optical studies \cite{menon1995bold}, which suggests that it is related to the specific experimental conditions such as object species, stimuli type, and measurement parameters etc.

\subsection{Co-Registered fMRI and fNIRS}

Although both fMRI and fNIRS targets the hemodynamic response, they are based on different physics principles and possess different advantages. Because fNIRS cannot provide anatomical references, co-registering with MRI is preferred, which helps in localizing the signal source. Figure 5 shows an example of a 3-D functional fNIRS mapping fused with MRI image. It is also suggested that fNIRS is best to be correlated with fMRI signals \cite{ferrari2012brief}. BOLD fMRI signal reflects dynamic changes of deoxy-hemoglobin concentration. However, this change may result from either an increase in oxygenation or a decrease in blood volume, which could be caused by various physiological conditions. By correlating fNIRS results with fMRI, it provides better calibration and leads to more accurate interpretations to the measurement.

\section{Conclusion}

We have briefly reviewed the fundamental principles of functional brain imaging by two photon microscopy and near infrared spectroscopy. Optical imaging technology has gone through dramatic development in recent years. Multiphoton fluorescent microscopy in vivo has led to many ground breaking understanding of neuron activities, while fNIRS has done remarkable accomplishments in clinical researches, especially on infants. Future development in instrumentation, data analysis and multimodality methods should open more opportunities for functional optical imaging, and provide promising research tools for both laboratory and clinical applications.

\cleardoublepage
\addcontentsline{toc}{section}{\listfigurename}
\listoffigures

\cleardoublepage

\end{document}